\documentclass[onecolumn,aps,unsortedaddress,showpacs,amsmath]{revtex4}

\begin{document}

\title{Probability density function of turbulent velocity fluctuation}

\author{Hideaki Mouri}
\affiliation{Meteorological Research Institute, Nagamine 1-1, Tsukuba 305-0052, Japan}
\email{hmouri@mri-jma.go.jp}

\author{Masanori Takaoka}
\affiliation{Department of Mechanical Engineering, Doshisha University, Kyotanabe, Kyoto 610-0321, Japan}
\email{mtakaoka@mail.doshisha.ac.jp}

\author{Akihiro Hori}
\altaffiliation[Affiliated with ]{Japan Weather Association.}

\author{Yoshihide Kawashima}
\altaffiliation[Affiliated with ]{Tsukuba Technology Research.}
\affiliation{Meteorological Research Institute, Nagamine 1-1, Tsukuba 305-0052, Japan}

\date{\today}

\begin{abstract}
The probability density function (PDF) of velocity fluctuations is studied experimentally for grid turbulence in a systematical manner. At small distances from the grid, where the turbulence is still developing, the PDF is sub-Gaussian. At intermediate distances, where the turbulence is fully developed, the PDF is Gaussian. At large distances, where the turbulence has decayed, the PDF is hyper-Gaussian. The Fourier transforms of the velocity fluctuations always have Gaussian PDFs. At intermediate distances from the grid, the Fourier transforms are statistically independent of each other. This is the necessary and sufficient condition for Gaussianity of the velocity fluctuations. At small and large distances, the Fourier transforms are dependent.
\end{abstract}

\pacs{47.27.Ak}

\maketitle

\section{INTRODUCTION}
\label{S1}

While velocity differences in turbulence $u(x+\delta x)-u(x)$ have attracted much interest, velocity fluctuations $u(x)$ themselves are also fundamental to describing the turbulence. Usually it is observed that the probability density function (PDF) of the velocity fluctuations is close to Gaussian \cite{B53,VC69}. However, some experiments yield a sub-Gaussian PDF \cite{NWLMF97,SD98,J98}, which has a less pronounced tail than a Gaussian PDF. The reason remains controversial because there have been no systematical studies.

The observed Gaussianity had been explained by applying the cental limit theorem to the Fourier transformation of the velocity fluctuations \cite{B53}. This theorem ensures that a sum of many independent random variables has a Gaussian PDF \cite{KS77}. However, owing to the steep energy spectrum of turbulence, the Fourier transforms have considerably different magnitudes. In this case, the central limit theorem is not applicable \cite{J98}. We first reconsider the condition for the velocity fluctuations to have a Gaussian PDF (Sec.~\ref{S2}).

For studies of velocity fluctuations, an experimental approach is preferable to the popular direct numerical simulation. Although the direct numerical simulation is useful in studying small-scale motions of turbulence, they are not of our interest because the velocity fluctuations are dominated by energy-containing large-scale motions. Those in the direct numerical simulation suffer from its artificial boundary condition, initial condition, and forcing. 

We obtain experimental data of velocity fluctuations in a wind tunnel downstream of a turbulence-generating grid (Sec.~\ref{S3}). The grid turbulence is not homogeneous in the mean-wind direction and thus allows us to study developing, fully developed, and decayed states of turbulence by increasing the distance between the anemometer and the grid. We find that the PDF of velocity fluctuations changes accordingly from a sub-Gaussian to Gaussian, and to hyper-Gaussian (Sec.~\ref{S4}). Here a hyper-Gaussian PDF has a more pronounced tail than a Gaussian PDF. We discuss the observed behavior by using the velocity fluctuations themselves, the velocity differences, and the Fourier transforms.

\section{THEORY FOR GAUSSIANITY}
\label{S2}

Suppose that velocity fluctuations $u(x)$ are measured in a turbulent flow repeatedly over the range $0 \le x < L$. The length $L$ is much greater than the correlation length $l_u  = \int \langle u(x+\delta x)u(x) \rangle {\rm d}\delta x / \langle u^2 \rangle$. Here $\langle \cdot \rangle$ denotes an ensemble average. The measurements serve as realizations of the turbulence. Each of them is expanded into a Fourier series as
\begin{equation}
\label{eq1}
u(x) = \sum_{n=1}^{\infty} a_n(u) \cos \left( \frac{2 \pi nx}{L} \right)
     +                     b_n(u) \sin \left( \frac{2 \pi nx}{L} \right).
\end{equation} 
The Fourier transforms $a_n(u)$ and $b_n(u)$ for $n \gg 1$ have Gaussian PDFs over the realizations \cite{BP01} (see also Ref. \cite{FB95} for a more mathematical explanation). This is because, for example, the Fourier transform $a_n(u)$ is obtained as 
\begin{eqnarray}
\label{eq2}
a_n(u) 
&=& \frac{2}{L} \int_{0}^{L} u(x) \cos \left( \frac{2 \pi n x}{L} \right) {\rm d}x 
\nonumber \\
&=& \frac{2}{L} \left( \int_{0}^{L/m}      ... \, {\rm d}x +
                       \int_{L/m}^{2L/m}   ... \, {\rm d}x + ... +
                       \int_{(m-1)L/m}^{L} ... \, {\rm d}x
                \right),
\end{eqnarray}
with $1 \ll m \le n$. The segment size $L/m$ is set to be large enough so that the correlation $\langle u(x+\delta x)u(x) \rangle$ has converged to zero at $\delta x = L/m$. The integrations $\int_{0}^{L/m} ...\, {\rm d}x$, $\int_{L/m}^{2L/m} ... \, {\rm d}x$, ..., and $\int_{(m-1)L/m}^{L} ... \, {\rm d}x$ are regarded as independent random variables of the same magnitude. Then $a_n(u)$ has a Gaussian PDF as a consequence of the central limit theorem. The variance $\langle a_n^2(u)\rangle$ is equal to the energy spectrum $E_n$ \footnote{
Note that the norm of the basis functions is not unity. If the norm were unity, the variance of the transform would be equal to $E_n/2$.}.

The above discussion is not applicable to the Fourier transforms $a_n(u)$ and $b_n(u)$ for $n \simeq 1$, i.e., those for large wavelengths. Nevertheless, as far as the wavelength is finite, it is possible to show Gaussianity of the corresponding Fourier transform by increasing the data length $L$ and hence the $n$ value. In the limit $L \rightarrow \infty$, the transforms $a_n(u)$ and $b_n(u)$ for $n \simeq 1$ become zero and do not contribute to the velocity fluctuations $u(x)$. We are able to assume safely that all the Fourier transforms have Gaussian PDFs. Since this fact is independent of detailed dynamics, it is universal.

If and only if all the Fourier transforms $a_n(u)$ and $b_n(u)$ are statistically independent of each other, the Gaussianity of them leads to Gaussianity of the velocity fluctuations $u(x)$. To demonstrate this, we use the characteristic functions $\phi_n^{(a)}(\tau)$ and $\phi_n^{(b)}(\tau)$ for the PDFs of $a_n(u) \cos (2\pi nx/L)$ and $b_n(u) \sin (2\pi nx/L)$ at any fixed spatial position $x$ \cite{KS77}:
\begin{subequations}
\label{eq3}
\begin{eqnarray}
\phi_n^{(a)}(\tau) &=&
\exp \left[ - \frac{\tau^2}{2!} E_n \cos^2 \left( \frac{2\pi nx}{L} \right) \right] \\
\phi_n^{(b)}(\tau) &=&
\exp \left[ - \frac{\tau^2}{2!} E_n \sin^2 \left( \frac{2\pi nx}{L} \right) \right].
\end{eqnarray}
\end{subequations}
Here $E_n \cos^2 (2\pi nx/L)$ and $E_n \sin^2 (2\pi nx/L)$ are the variances of $a_n(u) \cos (2\pi nx/L)$ and $b_n(u) \sin (2\pi nx/L)$. From the expansion formula (\ref{eq1}) and the independence of the Fourier transforms $a_n(u)$ and $b_n(u)$, it follows that the sum of the logarithms of the characteristic functions $\phi_n^{(a)}(\tau)$ and $\phi_n^{(b)}(\tau)$ is equal to the logarithm of the characteristic function $\phi(\tau)$ for the PDF of the velocity fluctuations $u(x)$: 
\begin{equation}
\label{eq4}
\ln \phi (\tau) = \sum^{\infty}_{n=1} 
                   \ln \phi _n^{(a)} (\tau) + \ln \phi _n^{(b)} (\tau)
                = - \frac{\tau^2}{2!} \sum^{\infty}_{n=1} E_n.
\end{equation}
Thus the velocity fluctuations $u(x)$ have a Gaussian PDF with the variance $\langle u(x)^2 \rangle = \textstyle \sum^{\infty}_{n=1} E_n$. The independence of the Fourier transforms also leads to the statistical independence of the velocity fluctuations $u(x)$ at different spatial positions.

Therefore, the necessary and sufficient condition for the velocity fluctuations to have a Gaussian PDF is the independence of the Fourier transforms. This is a good approximation for fully developed turbulence, where large-scale motions of energy-containing eddies are random and independent. Although turbulence contains small-scale coherent structures such as vortex tubes \cite{SA97}, their importance to the velocity fluctuations is negligible, i.e., being as small as the energy ratio of the dissipation range to the energy-containing range.

\section{EXPERIMENTS}
\label{S3}

The experiments were done in two wind tunnels of Meteorological Research Institute. Their test sections were of $0.8 \times 0.8 \times 3$ and $3 \times 2 \times 18\,{\rm m}$ in size (hereafter, respectively, the small and large tunnels). The small tunnel was used to study developing and fully developed states of grid turbulence, while the large tunnel was used to study fully developed and decayed states.

Turbulence was produced by placing a grid across the entrance to the test section. The grid consisted of two layers of uniformly spaced rods, the axes of which were perpendicular to each other. We used different grids in the small and large tunnels. The separation of the axes of their adjacent rods were 0.10 and 0.40\,m, respectively. The cross sections of the rods were $0.02 \times 0.02$ and $0.06 \times 0.06\,{\rm m}$, respectively. The mean wind was set to be $U \simeq 10\,{\rm m}\,{\rm s}^{-1}$ in the small tunnel and $U \simeq 20\,{\rm m}\,{\rm s}^{-1}$ in the large tunnel.

We simultaneously measured the streamwise ($U+u$) and transverse ($v$) velocities. They are velocity components that are parallel and perpendicular to the mean-wind direction, respectively. The measurements in the small tunnel were done on the tunnel axis from $d = 0.25$ to 2.00\,m downstream of the grid with an interval of 0.25\,m. Those in the large tunnel were done from $d = 3.00$ to 17.00\,m with an interval of 1.00\,m. The ranges of the measurement positions were to the limit of mechanical constraints of the wind tunnels. Since there was no overlap in the distance $d$ between the small- and large-tunnel measurements, the individual data are identified by their $d$ values.

We used a hot-wire anemometer, which was composed of a crossed-wire probe and a constant temperature system. The wires were 5\,$\mu$m in diameter, 1.25\,mm in effective length, 1.25\,mm in separation, and oriented at $\pm 45^{\circ}$ to the mean-wind direction. The wire temperature was 280$^{\circ}$C, while the air temperature was 29--30$^{\circ}$C in the small tunnel and 14--19$^{\circ}$C in the large tunnel. We calibrated the anemometer before and after the measurements. 

The signal was low-pass filtered with 24\,dB/octave and sampled digitally with 16-bit resolution. In the small-wind measurements, the filtering was at 8\,kHz and the sampling was at 16\,kHz. In the large-wind measurements, the filtering was at 20\,kHz and the sampling was at 40\,kHz. The entire length of the signal was as long as $5 \times 10^6$ points. We obtained longer data of $2 \times 10^7$ points at the positions $d = 0.25$, 2.00, 8.00, and 12.00\,m. 

The turbulence levels, i.e., the ratios of the root-mean-square values of the velocity fluctuations $\langle u^2 \rangle ^{1/2}$ and $\langle v^2 \rangle ^{1/2}$ to the mean streamwise velocity $U$, were always low ($\alt 0.2$; see below). This is a good characteristic of grid turbulence and allows us to rely on the frozen-eddy hypothesis of Taylor, $\partial / \partial t = -U \partial / \partial x$, which converts temporal variations into spatial variations in the mean-wind direction.

The resultant spatial resolution is comparable to the probe size, $\sim 1\,{\rm mm}$. Since the probe is larger than the Kolmogorov length, $0.1$--$0.2\,{\rm mm}$, the smallest-scale motions of the flow were filtered out. The present resolution is nevertheless typical of hot-wire anemometry \cite{VC69,SD98,FK67,CG99,AGHA84}.

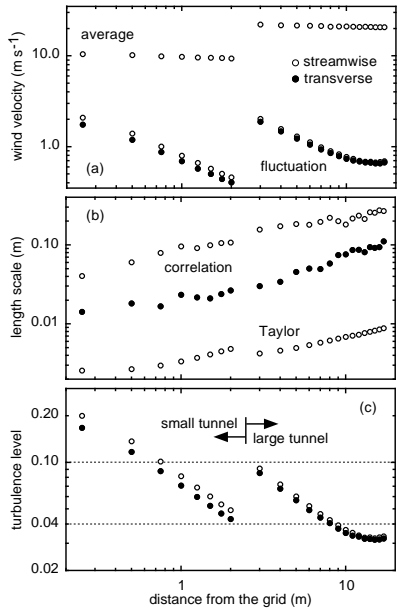
\begin{figure}[b]
\caption{\label{F1} (a) Mean streamwise velocity $U$ and root-mean-square values of velocity fluctuations $\langle u^2 \rangle^{1/2}$ and $\langle v^2 \rangle^{1/2}$. (b) Correlation lengths $l_u$ and $l_v$, and Taylor microscale $\lambda$. (c) Turbulence levels $\langle u^2 \rangle^{1/2}/U$ and $\langle v^2 \rangle^{1/2}/U$. The abscissa is the distance $d$ from the grid. The open circles denote the streamwise velocity $u$, while the filled circles denote the transverse velocity $v$. The horizontal dotted lines separate the turbulence levels for which the turbulence is developing, fully developed, and decayed. In the calculation of the Taylor microscale, the velocity derivatives were estimated as, e.g., $\partial u / \partial x$ = $[ 8 u(x+ {\mit\Delta} )- 8 u(x- {\mit\Delta} )-u(x+ 2 {\mit\Delta} )+u(x- 2 {\mit\Delta} )]/ 12 {\mit\Delta}$, where $\mit\Delta$ is the sampling interval.}
\end{figure}

Fig.~\ref{F1} shows the mean streamwise velocity $U$, the root-mean-square fluctuations $\langle u^2 \rangle ^{1/2}$ and $\langle v^2 \rangle ^{1/2}$, the correlation lengths $l_u$ and $l_v$, the Taylor microscale $\lambda = (\langle u^2 \rangle / \langle (\partial u/ \partial x)^2 \rangle )^{1/2}$, and the turbulence levels $\langle u^2 \rangle ^{1/2}/U$ and $\langle v^2 \rangle ^{1/2}/U$. For these and the other similar diagrams, open symbols denote the streamwise velocity while filled symbols denote the transverse velocity. The flow parameters change systematically with the distance from the grid, indicating a systematical change of the turbulence. This is especially the case in the turbulence levels. Our following studies suggest $\langle u^2 \rangle ^{1/2}/U \agt 0.1$ for the developing state, $\langle u^2 \rangle ^{1/2}/U \simeq 0.04$--0.1 for the fully developed state, and $\langle u^2 \rangle ^{1/2}/U \alt 0.04$ for the decayed state.

\section{RESULTS AND DISCUSSION}
\label{S4}

\subsection{Overview}

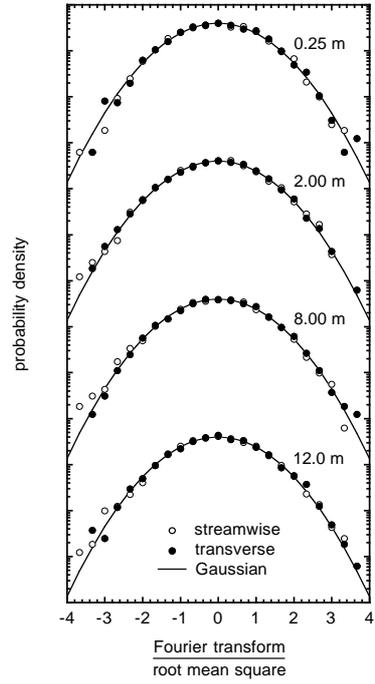
\begin{figure}[!]
\caption{\label{F2} PDFs of Fourier transforms of velocity fluctuations at the positions $d = 0.25$, 2.00, 8.00, and 12.00\,m. The wave number $k$ is 7\,m$^{-1}$. We vertically shift the PDFs by a factor of 10$^3$. The open circles denote the streamwise velocity $u$, while the filled circles denote the transverse velocity $v$. The solid lines denote Gaussian PDFs with zero mean and unity standard deviation. }
\end{figure}

Fig.~\ref{F2} demonstrates that Fourier transforms of the velocity fluctuations have Gaussian PDFs at the positions $d = 0.25$, 2.00, 8.00, and 12.00\,m. The individual data were divided into 4864 segments of 2$^{12}$ points, which were regarded as independent realizations of the turbulence. They were windowed by a flat-topped function, which rises from zero to unity in the first small fraction of the data and falls back to zero in the last small fraction. The PDFs shown in Fig.~\ref{F2} are those for the wave number $k = n/L = 7\,m^{-1}$, where the energy spectra at $d = 0.25$\,m have a peak (see below). Since the PDFs of $a_n$ and $b_n$ should be the same, they were put together in order to minimize statistical uncertainties. We also obtained Gaussian PDFs of the Fourier transforms at the other wave numbers and at the other positions in the wind tunnels.

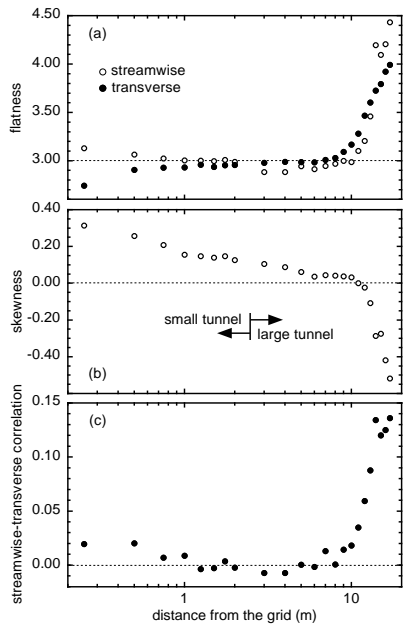
\begin{figure}[!]
\caption{\label{F3} PDFs of velocity fluctuations at the positions $d = 0.25$, 2.00, 8.00, and 12.00\,m. We vertically shift the PDFs by a factor of 10$^3$. The open circles denote the streamwise velocity $u$, while the filled circles denote the transverse velocity $v$. The solid lines denote Gaussian PDFs with zero mean and unity standard deviation.}
\end{figure}

However, velocity fluctuations do not necessarily have Gaussian PDFs. Fig.~\ref{F3} shows the PDFs at the positions $d = 0.25$, 2.00, 8.00, and 12.00\,m. The transverse-velocity PDF is sub-Gaussian at $d = 0.25$ m, Gaussian at $d = 2.00$ and 8.00\,m, and hyper-Gaussian at $d = 12.00$\,m. The streamwise-velocity PDF tends to be skewed owing to a shear flow. 


Fig.~\ref{F4} shows the flatness factors $F_u = \langle u^4 \rangle/\langle u^2 \rangle^2$ and $F_v$, the skewness factor $S_u = \langle u^3 \rangle / \langle u^2 \rangle^{3/2}$, and the streamwise-transverse correlation $C_{uv} = (\langle u^2 v^2 \rangle - \langle u^2 \rangle \langle v^2 \rangle) / [(\langle u^4 \rangle - \langle u^2 \rangle ^2) (\langle v^4 \rangle - \langle v^2 \rangle ^2)]^{1/2}$. With an increase of the distance from the grid, the transverse-velocity PDF changes from sub-Gaussian ($F < 3$) to Gaussian ($F = 3$), and to hyper-Gaussian ($F > 3$). The streamwise-velocity PDF tends to be skewed ($S \ne 0$) and changes from hyper-Gaussian to Gaussian, and to hyper-Gaussian. Also at large distances from the grid, the streamwise and transverse velocities have a significant correlation ($C_{uv} \gg 0$).


Since the streamwise fluctuations suffer from a shear, we are interested mainly in the transverse fluctuations. Their sub-Gaussian, Gaussian, and hyper-Gaussian PDFs are studied separately in the following subsections.

\subsection{Sub-Gaussian PDF in developing turbulence}

The transverse velocity has a sub-Gaussian PDF at the smallest distance from the grid, $d = 0.25\,{\rm m}$ (Figs.~\ref{F3} and \ref{F4}). Fig.~\ref{F5}(a) shows energy spectra of the streamwise and transverse velocities. They have peaks at the wave numbers $k \simeq 7$ and 14\,m$^{-1}$. The flow is in a transition state from quasi-periodic motions due to wavy wakes of the grid rods to weak turbulence. Fig.~\ref{F5}(b) shows the correlation coefficient between the Fourier transforms at adjacent wave numbers, $C_{nn'} = \langle a_n a_{n'} \rangle / (\langle a_n^2 \rangle \langle a_{n'}^2 \rangle)^{1/2}$ with $n' = n+1$. At around the energy peaks, we observe significant correlations. The Fourier transforms are not mutually independent.


The quasi-periodic motions should have finite correlation lengths. Namely, the motions should not be exactly periodic. If this were not the case, the central limit theorem would not be applicable to the calculation of the Fourier transforms (Sec.~\ref{S2}). Then the transforms would not have a Gaussian PDF. This is inconsistent with our result (Fig.~\ref{F2}).

For velocity differences $\delta u = u(x+\delta x)-u(x)$ and $\delta v = v(x+\delta x)-v(x)$, where $\delta x$ is the separation in the mean-wind direction, Fig.~\ref{F6} shows the variances $\langle \delta u^2 \rangle$ and $\langle \delta v^2 \rangle$, the flatness factors $F_{\delta u}$ and $F_{\delta v}$, the skewness factors $S_{\delta u}$ and $S_{\delta v}$, and the streamwise-transverse correlation $C_{\delta u \delta v}$. At small separations, we observe enhancements of $F_{\delta u}$, $F_{\delta v}$, $S_{\delta u}$ and $C_{\delta u \delta v}$. They are due to small-scale coherent structures \cite{NWLMF97,SA97,CG99,MOURI} and are not of our interest. At large separations, we observe oscillations of $\langle \delta u^2 \rangle$, $\langle \delta v^2 \rangle$, $F_{\delta u}$, $F_{\delta v}$, $S_{\delta u}$, $S_{\delta v}$, and $C_{\delta u \delta v}$. Their wavelengths roughly correspond to the wave number $k \simeq 7\,{\rm m}^{-1}$ of the spectral energy peak. The oscillations of $\langle \delta u^2 \rangle$ and $\langle \delta v^2 \rangle$ are in phase, while those of $S_{\delta u}$ and $S_{\delta u}$ are $\pm 90^{\circ}$ out of phase and those of $F_{\delta u}$, $F_{\delta v}$, and $C_{\delta u \delta v}$ are $180^{\circ}$ out of phase. 


Our results are explained if the velocity field is a superposition of a few quasi-periodic motions and a random background. It is actually possible to reproduce the oscillations and their phases qualitatively with a few sinusoidal functions satisfying the solenoidal condition and a random Gaussian noise. 

The quasi-periodic motions are predominant at local maxima of the variances $\langle \delta u^2 \rangle$ and $\langle \delta v^2 \rangle$. Since the flatness factor $F_{\delta v}$ is locally minimal and less than 3 there, the transverse-velocity amplitudes of those quasi-periodic motions lie in a limited range (see also Ref. \cite{J98}). This discussion would apply to the streamwise velocity as well. The oscillations of $S_{\delta u}$ and $S_{\delta v}$ imply the presence of quasi-periodic motions with different wave numbers that are coupled with each other (see also Fig.~\ref{F5}(b)). 

The background flow is predominant at local minima of the variances. Since the flatness factor $F_{\delta u}$ is greater than 3 there, the streamwise background tends to be intermittent. This tendency is not significant in the transverse background, which exhibits $F_{\delta v} \simeq 3$ even at its local maxima. The slight enhancement of $C_{\delta u \delta v}$ implies that the streamwise and transverse components of the background flow tends to have a correlation. 

Overall, the observed sub-Gaussianity of the transverse fluctuations is due to quasi-periodic motions that have finite amplitudes. The hyper-Gaussianity of the streamwise fluctuations is due to a background flow that tends to be intermittent.

\subsection{Gaussian PDF in fully developed turbulence}

The transverse velocity has a Gaussian PDF at intermediate distances from the grid, i.e., $d \agt 1\,{\rm m}$ in the small tunnel and $d \alt 10\,{\rm m}$ in the large tunnel (Figs. 3 and 4). This is because turbulence is fully developed there. Wavy wakes of the grid rods have evolved to energy-containing eddies that are random and independent. The corresponding Fourier transforms are thus independent. Actually we did not find spectral energy peaks such as those observed at $d = 0.25\,{\rm m}$. The wave-number--wave-number correlation in the energy-containing range $k \alt 10\,{\rm m}^{-1}$ is absent within the statistical uncertainty, $\vert C_{nn'} \vert \alt 0.1$.

The grid turbulence at the intermediate distances does not exhibit large-separation oscillations of $\langle \delta u^2 \rangle$, $\langle \delta v^2 \rangle$, $F_{\delta u}$, $F_{\delta v}$, $S_{\delta u}$, $S_{\delta v}$, and $C_{\delta u \delta v}$. The flatness factors $F_{\delta u}$ and $F_{\delta v}$ as well as the skewness factors $S_{\delta u}$ and $S_{\delta v}$ at large separations are identical to the Gaussian values. The streamwise-transverse correlation $C_{\delta u \delta v}$ at large separations is absent. These are observed usually in fully developed turbulence \cite{B53,VC69,FK67,MOURI}.

\subsection{Hyper-Gaussian PDF in decayed turbulence}

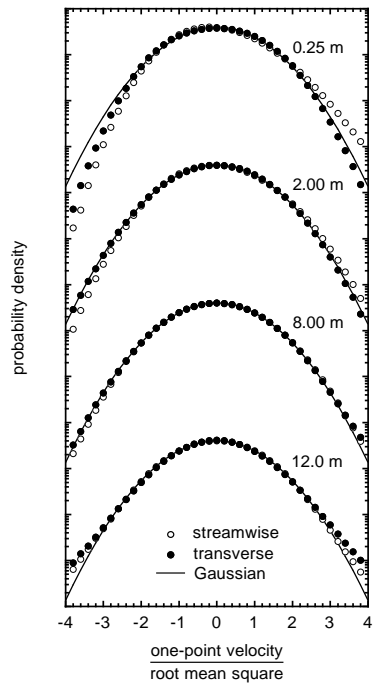
\begin{figure}[!]
\caption{\label{F4} (a) Flatness factors $F_u$ and $F_v$ of velocity fluctuations. (b) Skewness factor $S_u$. (c) Streamwise-transverse correlation $C_{uv}$. The abscissa is the distance $d$ from the grid. The open circles denote the streamwise velocity $u$, while the filled circles denote the transverse velocity $v$. The horizontal dotted lines indicate the values expected for independent Gaussian PDFs. The skewness factor for the transverse velocity $S_v$ is close to zero within the statistical error of about $\pm 0.02$.}
\end{figure}

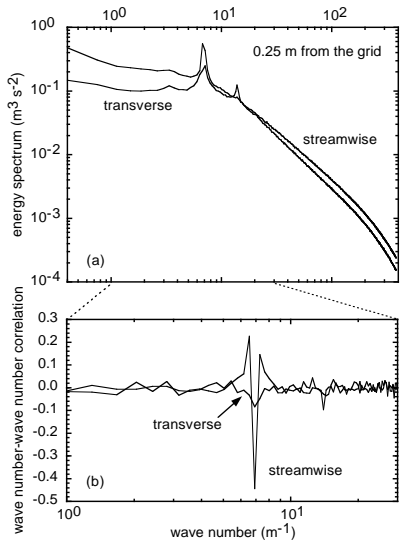
\begin{figure}[!]
\caption{\label{F5} (a) Energy spectra of velocity fluctuations at the position $d=0.25\,{\rm m}$. (b) Correlation between Fourier transforms at adjacent wave numbers $C_{nn'}$ with $n'=n+1$. The abscissa is the wave number $k = n/L$.}
\end{figure}

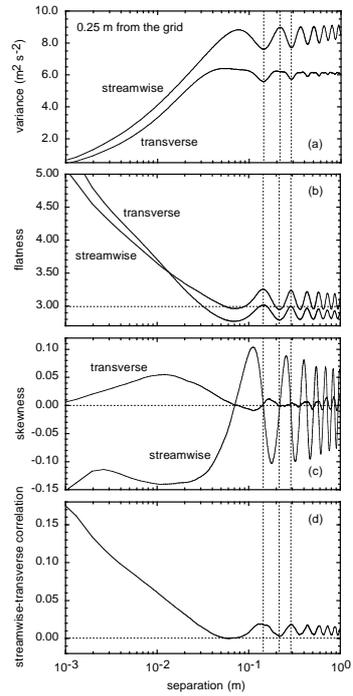
\begin{figure}[!]
\caption{\label{F6} (a) Variances $\langle \delta u^2 \rangle$ and $\langle \delta v^2 \rangle$ of velocity differences $\delta u$ and $\delta v$ at the position $d=0.25\,{\rm m}$. (b) Flatness factors $F_{\delta u}$ and $F_{\delta v}$. (c) Skewness factors $S_{\delta u}$ and $S_{\delta v}$. (d) Streamwise-transverse correlation $C_{\delta u \delta v}$. The abscissa is the separation $\delta x$. The horizontal dotted lines indicate the values expected for independent Gaussian PDFs.}
\end{figure}

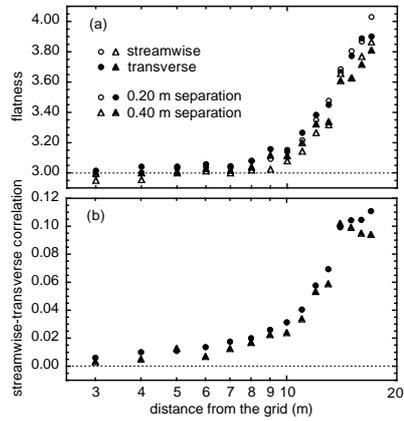
\begin{figure}[!]
\caption{\label{F7} (a) Flatness factors $F_{\delta u}$ and $F_{\delta v}$ of velocity differences $\delta u$ and $\delta v$ for separations $\delta x = 0.20\,{\rm m}$ (circles) and 0.40\,m (triangles). (b) Streamwise-transverse correlation $C_{\delta u \delta v}$. The abscissa is the distance $d$ from the grid. The open symbols denote the streamwise velocity $u$, while the filled symbols denote the transverse velocity $v$. The horizontal dotted lines indicate the values expected for independent Gaussian PDFs.}
\end{figure}

The streamwise and transverse velocities have hyper-Gaussian PDFs at largest distances from the grid, $d \agt 10\,{\rm m}$. The streamwise-transverse correlation $C_{uv}$ is also enhanced there (Figs.~\ref{F3} and \ref{F4}). For the separations $\delta x = 0.20\,{\rm m} \simeq l_u$ and $\delta x = 0.40\,{\rm m} \simeq 2l_u$, Fig.~\ref{F7} shows the flatness factors $F_{\delta u}$ and $F_{\delta v}$ and the streamwise-transverse correlation $C_{\delta u \delta v}$. They increase with the distance. Since the grid turbulence has decayed (Fig.~\ref{F1}), there remain only strong energy-containing eddies \cite{B53}, which intermittently pass the probe. The enhancement of the flatness factors $F_{\delta u}$ and $F_{\delta v}$ is due to enhanced values of $\delta u$ and $\delta v$, which should be associated with the surviving strong eddies. The streamwise-transverse correlation $C_{\delta u \delta v}$ is enhanced if $\delta u$ and $\delta v$ are enhanced simultaneously at the positions of those eddies. In this case, velocity fluctuations $u(x)$ and $v(x)$ are also enhanced at the eddy positions, resulting in the hyper-Gaussian PDFs as well as the significant streamwise-transverse correlation $C_{uv}$.


The presence of large-scale spatial structures implies a correlation among the corresponding Fourier transforms. Since the spatial structures consist of many Fourier modes, the correlation is not local in the wave number space. Although we failed to detect any significant correlation, we found moderate correlations with $\vert C_{nn'} \vert \simeq 0.2$ between many wave numbers in the energy-containing range $k \alt 10\,{\rm m}^{-1}$.

Roughly at the position where the flatness factors $F_u$ and $F_v$ begin to differ from the Gaussian value, the skewness factor for the streamwise velocity $S_u$ changes its sign (Fig.~\ref{F4}). This is probably because turbulence becomes weak. The positive skewness is more significant at a smaller distance from the grid and is attributable to the turbulence itself. On the other hand, the negative skewness is more significant at a larger distance. Using a low-pass filtering technique, we ascertained the presence of long-wavelength motions along the mean-flow direction ($k \ll 1\,{\rm m}^{-1}$). Their amplitude is larger and their PDF is more negatively skewed at a larger distance from the grid. These long-wavelength motions are attributable to effects of the wind tunnel, e.g., wall effects and deceleration of the mean flow.

\subsection{Small scale statistics}

Small-scale coherent structures such as vortex tubes in fully developed turbulence have attracted much interest \cite{NWLMF97,SA97,CG99,MOURI}. The statistics that are studied most often are the flatness factors $F_{\partial u/ \partial x}$ and $F_{\partial v/ \partial x}$ and the skewness factor $S_{\partial u/ \partial x}$ of the velocity derivatives $\partial u/ \partial x$ and $\partial v/ \partial x$. An increase of the microscale Reynolds number Re$_{\lambda} = \langle u^2 \rangle ^{1/2} \lambda / \nu$ is known to cause the increase of $F_{\partial u/ \partial x}$ and $F_{\partial v/ \partial x}$ and the decrease of $S_{\partial u/ \partial x}$ \cite{SA97}. Here $\nu$ is the kinematic viscosity. We briefly summarize overall trends of Re$_{\lambda}$, $F_{\partial u/ \partial x}$, $F_{\partial v/ \partial x}$, and $S_{\partial u/ \partial x}$ over distances from the grid. They are not of our interest but are expected to be helpful in a future experimental research.

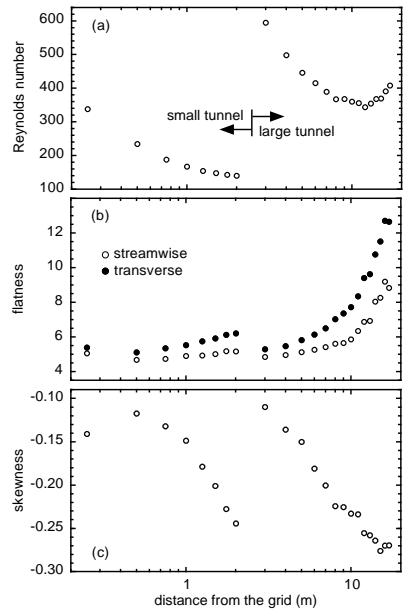
\begin{figure}[!]
\caption{\label{F8} (a) Reynolds number Re$_{\lambda}$. (b) Flatness factors $F_{\partial u/ \partial x}$ and $F_{\partial v/ \partial x}$ of velocity derivatives $\partial u/ \partial x$ and $\partial v/ \partial x$. (c) Skewness factor $S_{\partial u/ \partial x}$. The abscissa is the distance $d$ from the grid. The open circles denote the streamwise velocity $u$, while the filled circles denote the transverse velocity $v$. The skewness factor for the transverse velocity $S_{\partial v/ \partial x}$ is close to zero within the statistical error of about $\pm 0.01$.}
\end{figure}

The microscale Reynolds number Re$_{\lambda}$ decreases with the distance as shown in Fig.~\ref{F8}(a). The flatness factors $F_{\partial u/ \partial x}$ and $F_{\partial v/ \partial x}$ increase as shown in Fig.~\ref{F8}(b). The skewness factor $S_{\partial u/ \partial x}$ decreases as shown in Fig.~\ref{F8}(c). Thus the dependences of $F_{\partial u/ \partial x}$, $F_{\partial v/ \partial x}$, and $S_{\partial u/ \partial x}$ on Re$_{\lambda}$ are apparently opposite to those mentioned above. This is because the turbulence state changes from developing to fully developed, and to decayed with an increase of the distance. In order to compare with other experimental data, the measurement is required to be done at a position where the grid turbulence is in the fully developed state.

\section{CONCLUSION}
\label{S5}

The PDF of velocity fluctuations was studied systematically for grid turbulence. At small distances from the grid, where the turbulence is still developing, there are quasi-periodic motions having finite amplitudes, and hence the PDF is sub-Gaussian. At intermediate distances from the grid, where the turbulence is fully developed, motions of energy-containing eddies are random, and hence the PDF is Gaussian. At large distances from the grid, where the turbulence has decayed, there remain only strong eddies, and hence the PDF is hyper-Gaussian. The Fourier transforms of the velocity fluctuations always have Gaussian PDFs, in accordance with the central limit theorem. At intermediate distances from the grid, the Fourier transforms are statistically independent of each other. This is the necessary and sufficient condition for Gaussianity of the velocity fluctuations. At small and large distances, the Fourier transforms are dependent.

Our result serves as an example that the velocity fluctuations could have a sub-Gaussian PDF if there exist strong finite-amplitude motions. We suspect that this is the case in experiments of Noullez et al. \cite{NWLMF97} and Sreenivasan and Dhruva \cite{SD98}. They obtained $F_v \simeq 2.85$ in a free air jet and $F_u = 2.66$ in an atmospheric boundary layer, respectively (see also Refs. \cite{CG99,AGHA84}), for which no explanation has been proposed yet. The signal could suffer from finite-amplitude motions generated by the jet nozzle or the surface. It is of interest to analyze such data in the same manner as in our present work.

\begin{acknowledgments}
The authors are grateful to T. Umezawa for his help in our experiments and also to the referee for helpful comments.
\end{acknowledgments}



\begin{thebibliography}{999}

\bibitem{B53} G. K. Batchelor, {\it The Theory of Homogeneous Turbulence} (Cambridge University Press, Cambridge, U. K., 1953).

\bibitem{VC69} C. W. Van Atta and W. Y. Chen, J. Fluid Mech. {\bf 38}, 743 (1969).

\bibitem{NWLMF97} A. Noullez, G. Wallace, W. Lempert, R. B. Miles, and U. Frisch, J. Fluid Mech. {\bf 339}, 287 (1997).

\bibitem{SD98} K. R. Sreenivasan and B. Dhruva, Prog. Theor. Phys. Supple. Ser. {\bf 130}, 103 (1998).

\bibitem{J98} J. Jim\'enez, J. Fluid Mech. {\bf 376}, 139 (1998).

\bibitem{KS77} M. Kendall and A. Stuart, {\it The Advanced Theory of Statistics, vol. 1, 4th edition} (Griffin, London, U. K., 1977).

\bibitem{BP01} C. Brun and A. Pumir, Phys. Rev. E {\bf 63}, 056313 (2001).

\bibitem{FB95} Z. Fan and J. M. Bardeen, Phys. Rev. D {\bf 51}, 6714 (1995).

\bibitem{SA97} K. R. Sreenivasan and R. A. Antonia, Annu. Rev. Fluid Mech. {\bf 29}, 435 (1997).

\bibitem{FK67} F. N. Frenkiel and P. S. Klebanoff, Phys. Fluids {\bf 10}, 507 (1967).

\bibitem{AGHA84} F. Anselmet, Y. Gagne, E. J. Hopfinger, and R. A. Antonia, J. Fluid Mech. {\bf 140}, 63 (1984).

\bibitem{CG99} R. Camussi and G. Guj, Phys. Fluids {\bf 11}, 423 (1999).

\bibitem{MOURI} H. Mouri, H. Kubotani, T. Fujitani, H. Niino, and M. Takaoka, J. Fluid Mech. {\bf 389}, 229 (1999); H. Mouri, A. Hori, and Y. Kawashima, Phys. Lett. A {\bf 276}, 115 (2000).

\end{thebibliography}
\end{document}